\documentclass[final,numberedheadings,cmfonts]{aipproc}
\layoutstyle{6x9}
\usepackage{multirow}

\def\1{\mbox{l\hspace{-0.53em}1}}

\begin{document}

\title{The $[{\bf 70},1^-]$ baryon multiplet in the $1/N_c$ expansion revisited}


\author{N. Matagne and Fl. Stancu}{address={University of Li\`ege, Physics Department,\\
Institute of Physics, B5, \\ 
Sart Tilman, B-4000 Li\`ege 1, Belgium\\ 
E-mail: nmatagne@ulg.ac.be,
fstancu@ulg.ac.be}}

\begin{abstract}

The mass splittings of the baryons belonging to the $[{\bf 70},1^-]$-plet
are derived by using a simple group theoretical approach to the matrix elements 
of the mass formula. 
The basic conclusion is that
the first order correction to the baryon masses is of order 
$1/N_c$ instead of order $N^0_c$, as previously found. 
The conceptual difference between the ground state and the excited states is therefore removed.
\end{abstract}

\maketitle

{\it Introduction.}
The $1/N_c$ expansion of QCD  \cite{HOOFT}
is a powerful theoretical tool which allows to systematically analyze
baryon properties.
The success of the method stems from the discovery that the ground state baryons have 
an exact contracted SU(2$N_f$) symmetry 
when $N_c \rightarrow \infty $  \cite{Gervais:1983wq,DM},  $N_f$
being the number of flavors.
A considerable amount of work has been devoted to the ground state
baryons,
summarized in several review papers as, {\it e. g.}, 
\cite{DJM95,Jenkins:1998wy}.
For $N_c \rightarrow \infty $ the ground state baryons are degenerate. 
For large $N_c$ the mass splitting starts at order $1/N_c$.
The applicability of the approach to excited states is a subject of
current investigation. The  experimental facts indicate a small breaking
of SU($2N_f$) which make the $1/N_c$ studies of excited states plausible.
When the SU($N_f$) symmetry is exact, the baryon mass operator is 
a linear combination of terms 
\begin{equation}
\label{massoperator}
M  = \sum_{i} c_i O_i,
\end{equation} 
with the operators $O_i$  having the general form
\begin{equation}\label{OLFS}
O_i = \frac{1}{N^{n-1}_c} O^{(k)}_{\ell} \cdot O^{(k)}_{SF},
\end{equation}
where  $O^{(k)}_{\ell}$ is a $k$-rank tensor in SO(3) and  $O^{(k)}_{SF}$
a $k$-rank tensor in SU(2), but invariant in SU($N_f$).
The latter is expressed in terms of  SU($N_f$) generators. 
For the ground state one has $k$ = 0. The first factor gives the 
order $\mathcal{O}(1/N_c)$ of the operator in the series expansion.
The lower index $i$ represents a specific combination of generators, see
Table \ref{operators}. 
In  Eq. (\ref{massoperator}), each $O_i$ is multiplied 
by an unknown coefficient $c_i$ which is a reduced matrix element.
All these coefficients encode the QCD dynamics and are obtained from a fit to
the existing data. 
Additional terms are needed if SU($N_f$) is broken.

\vspace{0.3cm}


{\it Excited states.}
The excited states can be grouped into excitation bands 
with N = 1, 2, 3, etc. units of excitation energy.
Among these, the states belonging to the N = 1 band,
described by the $[{\bf 70},1^-]$-plet, 
have been most extensively studied,
either for $N_f$ = 2, see {\it e.g.} 
\cite{Goi97,CCGL,CaCa98}
or for  $N_f$ = 3 \cite{SGS}. 
The conclusion was that the splitting starts 
at order $N^0_c$.

The method has  been applied to the N = 2 band multiplets 
$[{\bf 56'},0^+]$ for $N_f$ = 2  \cite{CC00},  $[{\bf 56},2^+]$ 
for $N_f = 3$ \cite{GSS}  and  $[{\bf 70},\ell^+]$   for 
$N_f = 2$ \cite{MS2} and $N_f = 3$ \cite{Matagne:2006zf}
and  also to   $N_f = 3$
baryons \cite{MS1} of the  $[{\bf 56},4^+]$-plet
(N = 4 band).

The excited states belonging to  $[{\bf 56},\ell]$ multiplets are rather
simple and can be studied by analogy to the ground state. 
Naturally the splitting starts at order $1/N_c$ \ \cite{GSS,MS1}.

\vspace{0.2cm}
\begin{table}[ph]
\caption{List of operators and  coefficients of the mass operator
(\ref{massoperator}) for $[{\bf 70},1^-]$.}
\label{operators}
{\tiny
\renewcommand{\arraystretch}{2} 
\begin{tabular}{llrrrrr}
\hline
Operator &  & Fit 1 (MeV) &  Fit 2 (MeV) & Fit 3 (Mev) & Fit 4 (MeV) & Fit 5 (MeV) \\
\hline
$O_1 = N_c \ \1 $                                   &  $c_1 =  $  & $481 \pm5$  & $482\pm5$ &  $484\pm4$ &  $484\pm4$ & $498\pm3$\\
$O_2 = \frac{1}{N_c}\ell^i S^i$                     &  $c_2 =  $  & $-47 \pm39$ & $-30\pm34$ & $-31\pm20$ & $8\pm15$ & $38\pm34$\\
$O_3 = \frac{1}{N_c}S^iS^i$                         &  $c_3 =  $  & $161\pm 16$ & $149\pm11$ & $159\pm16$ & $149\pm11$ & $156\pm16$\\
$O_4 = \frac{1}{N_c}T^aT^a$                         &  $c_4 =  $  & $169\pm36$  & $170\pm36$ & $138\pm27$ & $142\pm27$ &\\
$O_5 = \frac{3}{N_c^2}\ell^{(2)ij}G^{ia}G^{ja}$     &  $c_5 =  $  & $-443\pm459$&            & $-371\pm456$&        & $-514\pm458$\\
$O_6 = \frac{1}{N_c^2}\ell^iT^aG^{ia}$              &  $c_6 =  $  & $473\pm355$ & $433\pm353$ &            &      & $-606\pm273$
\vspace{0.15cm} \\
\hline
$\chi_{\mathrm{dof}}^2$                             &             & $0.43$      & $0.68$ & $1.1$           & $0.96$ & $11.5$\\
\hline 
\end{tabular}}
\end{table}

The states belonging to $[{\bf 70}, \ell]$-plets are more 
difficult due to the presence of a mixed symmetry.  So far, in calculating 
the mass spectrum, the general practice 
was to split the baryon into an excited quark and a symmetric core,
in this way reducing  the problem to the well known ground state.
There are two drawbacks in this procedure.
One is that each generator of SU(2$N_f$) is written as a sum of two
terms, one acting on the excited quark and the other on the core. 
As a consequence, 
the number of linearly
independent operators to be used in Eq. (\ref{OLFS})
increases tremendously  and the number of coefficients 
to be determined becomes larger or much larger than the experimental data
available. For example, for the $[{\bf 70},1^-]$ multiplet with $N_f = 2$ 
one has 12 linearly independent
operators up to order $1/N_c$  \cite{CCGL}, instead of 6 
as in the present approach (see Table \ref{operators}). 
We recall that there are only 7 nonstrange 
resonances belonging to this band. They are given in Table \ref{MASSES}.
Consequently, in selecting the most dominant operators one has to make an
arbitrary choice  \cite{CCGL}.
The second drawback is  due to the 
truncation of  the wave function containing the
orbital and the spin-flavor parts. The exact wave function is
given by a linear combination of terms of equal weight where each term 
corresponds to a given Young tableau of mixed symmetry denoted 
by $[N_c-1,1]$.
In the above procedure
only the term where the last quark is in the second row
was kept, the other $N_c - 2$ terms being ignored. 
As a consequence the order of the spin-orbit operator became $N^0_c$.

In this practice
the matrix elements of the excited quark are straightforward, as being
described by single-particle operators. The matrix elements of the core operators 
$S^i_c$ and $T^a_c$ are also simple to calculate, while  $G^{ia}_c$
are more involved.
Analytic group theoretical formulas for several types of matrix elements of 
the SU(4) generators  have been derived in the late sixtieths  \cite{HP},
in the context of nuclear physics. 
Every matrix element is factorized 
according to a generalized Wigner-Eckart theorem into a reduced
matrix element and an SU(4) Clebsch-Gordan coefficient. 
Recently we have extended the approach to calculate isoscalar factors
needed for the matrix elements of
SU(6) generators between symmetric $[N_c]$ states\  \cite{Matagne:2006xx}.

\vspace{0.3cm}
{\it The $[\bf 70,1^-]$-plet.}
Here we propose a new method where the splitting into an excited quark and
a core is unnecessary. Details can be found in Ref. \ \cite{Matagne:2006dj}. 
All one needs to know are the matrix
elements of the SU(2$N_f$) generators between mixed symmetric
states   $[N_c-1,1]$. For $N_f$ = 2 these are provided by the 
work of Hecht and Pang \cite{HP}.
To our knowledge such matrix elements are yet 
unknown for  $N_f$ = 3. Thus our work deals with nonstrange 
baryon resonances only. 
The list of operators $O_i$ contributing to 
the mass operator (\ref{massoperator}) 
is shown in Table \ref{operators}
together with the coefficients $c_i$. The first column gives 
all linearly independent operators of type (\ref{OLFS}) up to order $1/N_c$.
The other columns indicate the values obtained for the dynamical 
coefficients $c_i$ from various fits. Fit 1 contains all operators and gives 
the best $\chi_{\mathrm{dof}}^2$. In the other columns one or two operators 
have been removed in order to understand their role in the fit. One can see 
that the removal of $O_5$ or of $O_6$ is not so dramatic but that
the removal of $O_4$ badly deteriorates the fit. The data are from 
Ref. \  \cite{Yao}.
It is important to note that here the angular momentum operator of components
${\ell_i}$ ($i$ = 1,2,3) is an intrinsic operator, acting on the entire 
system, not  on the $N_c$-th quark only, as in previous studies,
which means it was taken relative to a fixed center of mass
\ \cite{CCGL}. In other words here we do not have a center of mass problem
which may complicate the $N_c$ counting. 
\begin{center}
\begin{table}[h!]
\caption{Partial contributions and total mass (MeV) 
of $[\bf 70,1^-]$ resonances predicted by the 
$1/N_c$ expansion. 
The last two columns reproduce the experimental  masses and the status 
of resonances \cite{Yao}.}
\label{MASSES}
\renewcommand{\arraystretch}{2.5}
{\tiny
\begin{tabular}{crrrrrrcccl}\hline 
                    &      \multicolumn{6}{c}{Part. contrib. (MeV)}  & \hspace{.0cm} Total (MeV)   & \hspace{.0cm}  Exp. (MeV)\hspace{0.0cm}& &\hspace{0.cm}  Name, status \hspace{.0cm} \\

\cline{2-7}
                    &   \hspace{.0cm}   $c_1O_1$  & \hspace{.0cm}  $c_2O_2$ & \hspace{.0cm}$c_3O_3$ &\hspace{.0cm}  $c_4O_4$ &\hspace{.0cm}  $c_5O_5$ & $c_6O_6$   &        \\
\hline
$^2N_{\frac{1}{2}}$        & 1444 & -16 & 40 & 42 & 0 & -13  &   $1529\pm 11$  & $1538\pm18$ & & $S_{11}(1535)$****  \\
$^4N_{\frac{1}{2}}$        & 1444 &  39 & 201& 42 & -31& -33 &   $1663\pm 20$  & $1660\pm20$ & & $S_{11}(1650)$**** \\
$^2N_{\frac{3}{2}}$        & 1444 & -8  & 40 & 42 & 0  &  7  &   $1525\pm 8$   & $1523\pm8$  & & $D_{13}(1520)$****\\
$^4N_{\frac{3}{2}}$        & 1444 & 16  & 201& 42 & 25 & -13 &   $1714\pm45$   & $1700\pm50$ & & $D_{13}(1700)$***\\
$^4N_{\frac{5}{2}}$        & 1444 & -24 & 201& 42  & -6 & 20 &   $1677\pm8$    & $1678\pm8$  & & $D_{15}(1675)$****\\
\hline
$^2\Delta_{\frac{1}{2}}$  &  1444 & 16  & 40 & 211 & 0  & -66   & $1645\pm30$  & $1645\pm30$ & & $S_{31}(1620)$**** \\
$^2\Delta_{\frac{3}{2}}$  &  1444 & -8  & 40 & 211 & 0  & -33   & $1720\pm50$  & $1720\pm50$ & & $D_{33}(1700)$**** \vspace{0.15cm} \\
\hline 
\end{tabular}}
\end{table}
\end{center}

The partial contributions and the total mass predicted by the 
$1/N_c$ expansion are given in Table 2.
One can see that the contributions of all terms containing
angular momentum, and in particular that of the spin-orbit,   
are small. This can be viewed as a dynamical effect. An entirely new 
quantitative result is that the isospin-isospin 
term $O_4$ brings a dominant contribution to $\Delta$ resonances, of the
same order as the spin-spin term $O_3$ brings to $N$ resonances.  

\vspace{0.3cm}
{\it Conclusions.} 
These results shed a new light into the description 
of the baryon multiplet  $[{\bf 70},1^-]$ in the  $1/N_c$ expansion. 
The main findings are:
\begin{itemize}
 \item In the mass formula the expansion starts at order $1/N_c$, as for the ground 
 state, instead of $N^0_c$ as previously concluded.
 \item The isospin operator  $O_4 = \frac{1}{N_c} T^aT^a$ is crucial in the
 fit to the existing data and its contribution is as important as that 
 of the spin term $O_3 = \frac{1}{N_c} S^iS^i$. 
 \end{itemize}
 It would be interesting to reconsider the study of higher excited 
  baryons, for example those belonging to $[{\bf 70},\ell^+]$ multiplets, 
  in the spirit of the present approach.
  Based on group theoretical  arguments it is expected that the mass 
  splitting starts at order $1/N_c$, as a general rule.


\end{document}